\documentstyle[epsfig]{elsart}

\begin{document}
 
 \begin{frontmatter}
\title{Mobile Bipolarons in the Adiabatic Holstein-Hubbard Model 
in 1 and 2 dimensions.}
\author{L. Proville  \thanksref{email1} and S. Aubry 
\thanksref{email2}}
\address{Laboratoire L\' eon Brillouin (CEA-CNRS), CE Saclay\\
	91191-Gif-sur-Yvette Cedex, France}
\thanks[email1]{email: proville@bali.saclay.cea.fr}
\thanks[email2]{email: aubry@bali.saclay.cea.fr}

\begin{abstract}
	The bound states of two electrons in the adiabatic 
Holstein-Hubbard model are studied numerically in one and two
dimensions from the anticontinuous limit. This model involves a
competition between a local electron-phonon coupling (with a classical 
lattice) which tends to form
pairs of electrons and the repulsive Hubbard interaction $U \geq 0$ which
tends to break them. 

	In 1D, the ground-state always consists in a pair of localized polarons
	in a singlet state. They are located at the same site for $U=0$.
	Increasing U, there is a first order transition at which
	 the bipolaron becomes a spin singlet pair of two  polarons bounded by 
	a magnetic interaction.
	The pinning mode of the bipolaron soften in the vicinity of this 
	transition leading to a higher mobility of the bipolaron
	which is tested numerically.

	In 2D, and for any $U$, the electron-phonon coupling needs to be
	large enough in order to form small polarons or bipolarons instead of
	extended electrons.
	We calculate the phase diagram of the bipolaron involving 
	first order transitions lines with a triple point. 
	 A pair of polarons can form  three types of bipolarons:
	 a) on a single site at small $U$, b) a spin singlet state on two nearest
	 neighbor sites for larger $U$ as in 1D and  c) a new
	 intermediate state obtained as the resonant combination of four
	 2-sites singlet states sharing a central site, called quadrisinglet.

	 The breathing and pinning internal modes of bipolarons in 2D
	 generally only weakly soften and thus, they are practically not mobile.
	 On the opposite, in the vicinity of the
	 triple point involving the quadrisinglet,
	 both modes exhibit a significant softening. However, it 
	 was not sufficient for allowing the existence of a classical mobile 
	 bipolaron (at least in that model).

\end{abstract}
\end{frontmatter}

\section{Introduction}
It is well-known for many decades since \cite{Lan33} that a single electron
submitted to an electron phonon coupling may generate a polaron that is
a quasi-particle consisting into an electron localized in a self
consistent lattice  potential. For a local electron phonon interaction,
the formation of a polaron always occurs in 1D. At small coupling, this
polaron becomes large and highly mobile. On the opposite, in 2D (and more
dimensions), the formation of a polaron requires a large enough coupling
\cite{Emi82}. Then, the polarons are always small that is mostly localized
on a single site and not mobile.
 
If there is no electron repulsion, two polarons form a bipolaron with
 two electrons in a spin singlet state localized in the same potential 
 well. When there is a strong enough electronic repulsion, the bipolaron
 may be broken into two unbound polarons. However, in the intermediate regime, 
 we are going to show that there are new bounded bipolaronic states with 
 interesting properties.
 
There have been many tentative theories about bipolaronic superconductivity for many years (see \cite{Rev95}  for a review). In the well-known BCS
theory valid at weak electron phonon coupling, the electron pairs
(Cooper pairs) form self
consistently in the superconducting phase only and are spatially very 
extended.  For the bipolaronic 
superconductivity, it is speculated that when there are pre-existent 
bounded pairs of electrons (small bipolaron), that they may behave
as a quantum boson liquid and condense into a superfluid state. 
However, the most serious critiscismes to these theories is that when the 
bipolarons exist in 2 and 3 D models (with realistic physical parameters),
their effective mass is so huge that their quantum character completely 
disappear. They should condense into spatially ordered or 
disordered structures which could be just considered as a chemical bond 
ordering. 

Expansions at large electron phonon coupling \cite{AAR86},
modelize the
bipolaronic system as an array of coupled quantum spins $1/2$ with 
two types of coupling. The spin is $+1/2$ at a site 
occupied by a bipolaron and $-1/2$ in the opposite case. There is an 
$x-y$ coupling between neighboring spins describing the bipolaron 
tunnelling and which yields the 
bipolaron band width and a $z-z$ coupling representing the potential 
interaction between the bipolarons. 
If the $x-y$ term dominates, the spin ordering occurs in the plane $x-y$
and the structure is superconducting. If the $z$ term dominates,
one get a spatial ordering of the pseudospin in the $z$ direction that
is a spatial ordering of the bipolarons (or chemical bonds) in the real 
space. The real systems where the 
effective mass of the electrons are usually much smaller than the atomic 
masses are  generally close to the adiabatic limit. In that case the $x-y$
couplings turns  out to be negligible compared to the $z$ interaction
which eliminates any possibility of superconductivity but instead of,
 a spatial bipolaron
ordering. It is clear that this regime is quite well-described within 
a purely adiabatic approximation (the {\it chemist} approximation). 

However, it might exist special situations where this $x-y$ coupling 
could be  enhanced 
to relative large values and dominates the $z$ coupling
thus favoring a bipolaronic superconductivity at unusually high temperatures. 
The purpose of this paper is to show
that in a restricted region of the parameter space only,
the competition between the electron phonon coupling and the
electron-electron repulsion could produce a sharp increase in the mobility of
bipolarons while simultaneously the binding energy of the bipolaron remains
significantly large thus favoring superconductivity as conjectured 
 in \cite{Aub92,Aub95}.
 
We test these ideas on the simplest model which is 
 the Holstein Hubbard model, where both electron-phonon 
interactions and electron-electron interaction are involved.  In a first step,
we consider the adiabatic 
 model where the quantum fluctuations of the lattice are neglected
 and shall look at the bipolaron classical mobility only. The next step will
 be done in a forthcoming work, 
where it will be shown that when the bipolarons are very mobile,
the adiabatic approximation is not valid because they are very
sensitive to the quantum lattice fluctuation by producing a sharp increase
of the  bipolaron bandwidth.  
     
\section{The Model: definitions}
In order to make the physical parameters explicit, we write the
 Holstein-Hubbard Hamiltonian with its full set of parameters:

\begin{equation}
{\cal H} = -T\sum_{<i,j>,\sigma}C_{i,\sigma}^{+}C_{j,\sigma}
+\sum_{i}  \hbar \omega_0(a_{i}^{+}a_{i})+g n_{i}(a^{+}_{i}+a_i)
+{\upsilon}n_{i,\uparrow}n_{i,\downarrow}   \label{hamiltonian}
\end{equation}

where $T$ is the transfer integral between nearest neighbor sites $<i,j>$ 
of the lattice, of the electrons represented by the standard fermion 
operators $C_{i,\sigma}^{+}$  and $C_{j,\sigma}$ at site $i$ with 
spin $\sigma = \uparrow$ or $\downarrow$.
$a^{+}_i$ and $a_i$ are standard creation and annihilation boson 
operators of phonons. 
$\hbar \omega_0$ is the phonon energy of a dispersionless optical phonon 
branch. $g$ is the constant of the onsite electron
phonon coupling. The onsite electron-electron 
interaction is represented by a Hubbard term with positive coupling 
${\upsilon}$. 

Choosing the energy $E_0=8 g^2/\hbar\omega_0$ as energy unit
and introducing the position and momentum operators:

\begin{eqnarray}
&u_i&=\frac{\hbar\omega_0}{4g}(a_{i}^{+}+a_{i}) \label{position}\\
&p_{i}&=i\frac{2g}{\hbar\omega_0}(a_{i}^{+}-a_{i}) \label{momentum}
\end{eqnarray}

we obtain the dimensionless hamiltonian:

\begin{equation}
H = \sum_i{\left(\frac{1}{2} u_i^2 + \frac{1}{2}u_i n_i
+U n_{i\uparrow}n_{i\downarrow} \right)}
-\frac{t}{2}\sum_{<i,j>,\sigma} C_{i,\sigma}^{+}C_{j,\sigma} 
+ \frac{\gamma}{2} \sum_{i}p_i^2 \label{hamiltreduc}
\end{equation}

The parameters of the system are now:
\begin{equation}
   E_0=8 g^2/\hbar\omega_0 \qquad   U = \frac{\upsilon}{E_{0}} \qquad 
	t =  \frac{T}{E_{0}} \qquad 
	 \gamma = \frac{1}{4} (\frac{\hbar\omega_0}{2g})^4  
	\label{param} 
\end{equation}


As soon as the electron phonon coupling $g$ becomes reasonably large
 that is larger than the phonon energy $\hbar \omega_0$, $\gamma$ becomes
 very small. The adiabatic approximation is obtained by taking $\gamma=0$.
 We shall assume this condition from nowon. 
   
Then $\{u_i\}$ commutes with the Hamiltonian and can be taken as a scalar 
variable. For a given set of $\{u_i\}$, the 2-electron ground-state of
the Hamiltonian 

\begin{equation}
H_{ad} = \sum_i{\left(\frac{1}{2} u_i^2 + \frac{1}{2}u_i n_i+U n_{i\uparrow}
n_{i\downarrow}\right)}
-\frac{t}{2}\sum_{<i,j>,\sigma}C_{i,\sigma}^{+}C_{j,\sigma} 
\label{hamiltelectr}
\end{equation}

has to be searched among singlet states with the form

\begin{equation}
|\Psi>=\sum_{i,j}\psi_{i,j}C_{i,\uparrow}^{+}C_{j,\downarrow}^{+}|\emptyset>
\label{Psi_el}
\end{equation}

where $\psi_{i,j}=\psi_{j,i}$ is normalized $\sum_{i,j} |\psi_{i,j}|^2 =1$.
Then the energy of the system depends on $|\psi>=\{\psi_{i,j}\}$ and
$\{u_i\}$ as

\begin{equation} \label{eq3}
F(\{\psi_{i,j}\},\{u_i\}) = \sum_i{\left(\frac{1}{2} u_i^2 +  u_i \rho_i
+ U |\psi_{i,i}|^2\right)} - \frac{t}{2} <\psi|\Delta|\psi>
\end{equation}

where 
\begin{equation}
\rho_i=\frac{1}{2}\sum_j (|\psi_{i,j}|^2+|\psi_{j,i}|^2)
\label{rho_i}
\end{equation}

 is the half electronic density at site $i$ and $\Delta$ is a discrete
 Laplacian operator in $2d$ dimensions ($d$ being the initial
 lattice dimension).
 
 \section{The anticontinuous limit}
 
 The minimization of $F(\{\psi_{i,j}\},\{u_i\})$ with respect to 
 the normalized electronic state $\{\psi_i^*\}$ yields the electronic 
 eigenequation 
 
 \begin{equation}
	- \frac{t}{2} \Delta \psi_{i,j} +
	\left(\frac{1}{2} (u_i+u_j) + U \delta_{i,j}\right) \psi_{i,j} =
	F_{el}(\{u_i\})  \psi_{i,j}
	\label{eleigen}
\end{equation}

where $F_{el}$ is the electronic ground-state energy in the  
potential generated by $\{u_i\}$. Then for $t > 0$, the wave function $\psi_{i,j}$ 
has to be real, positive and symmetric. The adiabatic  potential for
the atoms is then

\begin{equation}
F_{ad}(\{u_i\}) = \sum_i \frac{1}{2} u_i^2  + F_{el}(\{u_i\})
\label{adpot}
\end{equation}
It has infinitely many minima \cite{AAR92,Aub93,BMK94} close to
the anticontinuous limit $t=0$. An adiabatic configuration $\{u_i\}$
is metastable when it is a local minima of \ref{adpot}  \cite{AAR92}. 

The minimization of $F(\{\psi_{i,j}\},\{u_i\})$ can also be done 
first with respect to $\{u_i\}$ which yields $u_i+\rho_i=0$. Substitution
in \ref{eq3}, yields a different variational form on $\{\psi_{i,j}\}$
non equivalent to \ref{adpot}

\begin{equation}
\label{eq3bis}
F_{\psi}(\{\psi_{i,j}\},\{u_i\}) = \sum_i{\left(-\frac{1}{2} \rho_i^2
+ U \sum_i |\psi_{i,i}|^2\right)} - \frac{t}{2} <\psi|\Delta|\psi>
\end{equation}

The extremalization of \ref{eq3bis} with respect to $\psi_{i,j}$
with the condition of normalization, 
yields a generalized discrete nonlinear Schroedinger equation
on a $2d$-dimensional lattice 

\begin{equation}
	- \frac{t}{2} \Delta \psi_{i,j} +
	\left(-\frac{1}{2} (\rho_i+\rho_j) + U \delta_{i,j}\right) \psi_{i,j} =
	F_{el}  \psi_{i,j}
	\label{NLGS}
\end{equation}
corresponding to \ref{eleigen}. 

This equation \ref{NLGS} has also an anticontinuous limit at $t=0$
which is different of those of model \ref{adpot}. It yields more states 
at this limit because it is not required that the electronic eigen states 
be in its ground-state with respect to the lattice potential.  This allows 
one to involve electronic excitations  but it  has of course the same
ground-states. The classification of these states will not be discussed here.

However for eq.\ref{NLGS} at $t=0$, the phases of each complex number $\psi_{i,j}$ is 
arbitrary. This situation is analogous to the anticontinuous limit of the breather
problem  where the  phase of the uncoupled oscillators is degenerate (see 
ref.\cite{MA94,Aub97} for details).  The consequence is that the implicit
function theorem cannot be applied directly for proving the possible continuation
of the solutions at $t=0$. For the breather problem, a trick has been to consider
first only  time reversible solutions thus removing this phase degeneracy 
but this was not absolutely necessary \cite{Aub97,SMK97}). In our model, it is
also convenient to consider in a first step only real electronic wave functions.
Then, these real normalized solutions at $t=0$ can be
continued for $t \neq 0$ not too large. This is not a restriction for 
finding the electronic ground-states because we know that the 
corresponding electronic state is real, positive and symmetric.  

\begin{figure} \begin{center}
\includegraphics{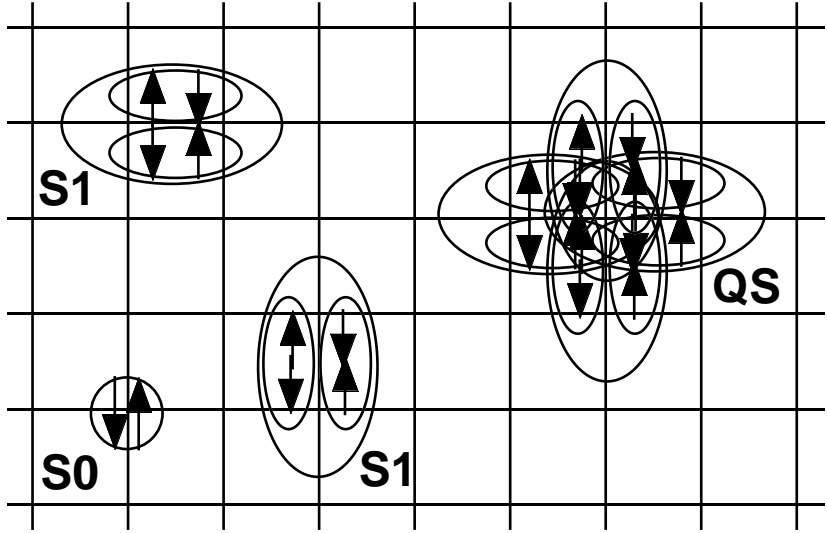}
\caption{\label{fig1} Schemes representing on a 2d lattice 
a) a standard onsite  bipolaron (S0)  b) two-site singlet bipolarons ("Spin Resonant 
Bipolaron") (S1) in $x$ and $y$ directions 
c) a quadrisinglet bipolaron localized on four bonds (QS).}
\end{center}\end{figure}

This continuation can be done by numerical methods. 
Most of the real solutions of \ref{NLGS} are unstable for $t$ small but some of them
can recover stability at larger $t$ and even become the ground-state.
We calculated numerically with a high accuracy a few number of these solutions
rather well localized on a small number of sites (also taking advantage of
its spatial symmetries if any) which we believed to be possible candidate for
being a ground-state. By comparing their energies one with each others, 
the bipolarons state are found which are presumably the exact ground-state. 
In the domain of parameter we explored, we found mostly three kinds of
bipolaronic states with a significantly large binding energy (see fig.\ref{fig1}) as
ground-states in some domain of $U$ and $t$. \footnote{However, there 
are unexplored domains for large $U$ where the ground-states will be 
different but their very weak binding energy makes that they are less 
interesting at the present stage of our approach.} They are
obtained by continuation from the following solutions of 
eq. \ref{NLGS}  at $t=0$: 

1-the standard onsite bipolaron denoted (S0) at given site $i$ where

\begin{equation}
	\psi_{i,i} = 1  \quad\mbox{and}\quad \psi_{m,n} = 0 \quad \mbox{else for}
	\quad (m,n) \neq (i,i)
	\label{bip0}
\end{equation}

2- the two-site singlet  bipolaron , localized on two nearest neighbor 
sites $i$ and $j$ denoted (S1) which was named  "Spin Resonant" bipolaron
in ref.\cite{Aub92,Aub95}. We also observed ground-states which are singlet states
where the polarons  are at distance $2$ (denoted (S2) but with a weaker 
binding energy.

\begin{eqnarray}
	\psi_{i,j} &=& \psi_{j,i}=\frac{1}{\sqrt{2}} 
	\quad \mbox{and}\nonumber\\
	\psi_{m,n} &=& 0 \quad \mbox{else for}
	\quad (m,n) \neq (i,j) \quad \mbox{and} \quad (m,n) \neq (j,i)
	\label{bip1}
\end{eqnarray}

3-and in two dimensions, a new unexpected solution which is
the quadrisinglet localized bipolaron
denoted (QS) which is localized at the given  site $i$
and its four nearest neighbors $j_{\nu}$ ($\nu=1,..,4$).
It is the linear combination of four singlets located on the four 
nearest neighbor bonds $<i,j_{\nu}>$

\begin{eqnarray}
	\psi_{i,j_{\nu}} &=& \psi_{j_{\nu},i}=\frac{1}{\sqrt{8}} 
	\quad \mbox{and} \nonumber\\
	\psi_{m,n} &=& 0 \quad \mbox{else for}
	\quad (m,n) \neq (i,j_{\nu}) \quad \mbox{and} \quad (m,n) \neq (j_{\nu},i)
	\label{bip0p}
\end{eqnarray}

This solution can be viewed as a localized RVB state similar to those 
proposed by Anderson some years ago \cite{And87} in the pure Hubbard model
in 2D as a theory for superconductivity in cuprates.
The physical origin of the binding of the bipolaron in states (S1), 
(S2).. and (QS) can be interpreted  of magnetic origin. When the 
Hubbard term increases too much, bipolaron (S0) could break into two polarons 
far apart but then both of them are magnetic with a spin $1/2$. They should
interact by an antiferromagnetic exchange coupling, as predicted by standard
perturbation theories. Then, for moderately large Hubbard terms, it 
remains more favorable to reduce the distance between
the two polarons for gaining a magnetic energy by forming a
singlet state (S1), (S2)\ldots.  
We have no simple interpretation to explain why bipolaron (QS) which 
has a more complex structure and could become the most stable in 2D models
but is not in 1D (as far in our domains of investigation).

\section{Bipolaronic Ground-states in One dimension}

\begin{figure}\begin{center}
\includegraphics [width=12cm] {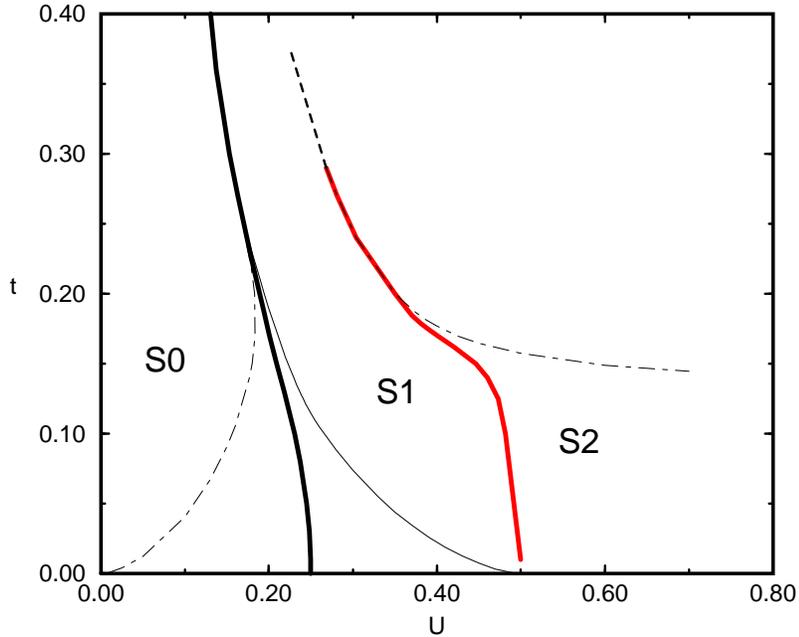}
	\caption{\label{fig2}
Phase diagram versus $U$ and $t$ of the bipolaronic ground-state.
showing first order transition lines (solid lines) between bipolarons (S0) and (S1) 
and between bipolarons (S1) and (S2).
The thin line is the limit of  metastability of bipolaron  (S0) 
and the two dot-dashed lines are those of (S1).}
\end{center}\end{figure}

The size of the system was chosen large
enough compared to the bipolaron size (in our calculations up to the size 41
for spatially symmetric states).
As expected, we found that for any value of $t$ (tested up to $t=0.6$), 
there is always a biplaronic state which is lower in energy than 
a pair of extended electrons. The phase diagram is shown fig.\ref{fig2}.

The first order transition lines are sharp and well defined for $t$ small
when the bipolarons have a small spatial extension.  
When $t$ increases, the size of the bipolarons also increases and
it becomes practically impossible to distinguish numerically the energy
differences between bipolaronic states obtained by continuation from  
different states at the anticontinuous limit.
This can be  interpreted by the fact that when $t$ is large, such a 
model is well described with a continuous space variable $x$ instead of the 
discrete lattice site $i$. This continuous model exhibit only one bipolaronic
solution which "erases" all the transitions due to discreteness.

\begin{figure}\begin{center}
\includegraphics [width=12cm] {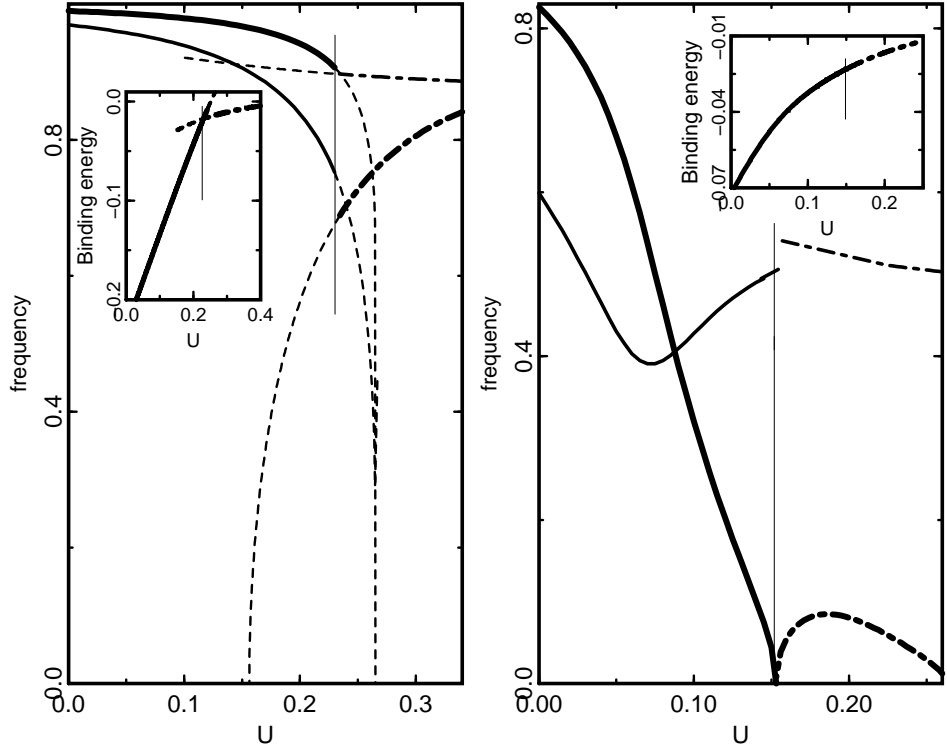}
\caption{\label{fig3}
Frequencies versus $U$ of the pinning modes (thick lines) and the breathing 
modes (thin lines) for bipolarons (S0) (full lines) and (S1) (dot-dashed 
lines) at $t=0.1$ (left) and  $t=0.3$ (right).
The inserts show the corresponding binding energies
for breaking the bipolarons into two polarons far apart.}
\end{center}\end{figure}

We now look at the possible mobility of the obtained bipolarons.
The precise calculation of the Peierls-Nabarow(PN) energy barrier requires too much
numerical work while the calculation of the pinning modes of 
the bipolarons  is much easier and brings nevertheless a similar information.
The eigenvalues of the matrix of the second variation of the lattice potential
energy $F_{ad}(\{u_i\})$ \ref{adpot} are the squares of the phonon frequencies of the
bipolaron calculated within the standard Born-Oppenheimer approximation
 (in unit $\sqrt{\gamma}$). Beside the initial flat optical branch of phonon 
 at frequency $1$, there are two localized mode. One is spatially 
 antisymmetric and correspond to the pinning mode of the bipolaron. The 
 second one is spatially symmetric and is a breathing mode.  
There is a  phonon  softening at the first order transition between 
bipolaron (S0) and (S1) which becomes almost complete for $t$ large 
enough. Fig.\ref{fig3} shows for $t=0.3$, that 
the frequency of the bipolaron (S0) practically vanishes at
the transition with (S1) which becomes almost second order. 

\begin{figure} \begin{center}
\includegraphics[angle=270,width=12cm]{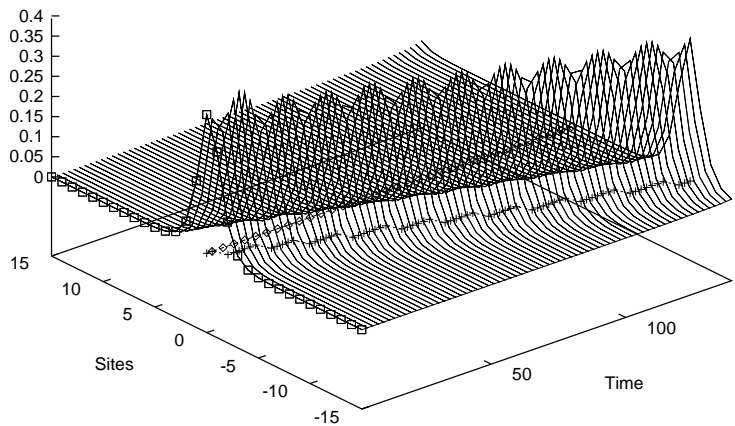}
\caption{\label{fig4}
3D Plot of the electronic density $\{\rho_i\}$ versus 
time for a moving bipolaron (S0) at $t=0.3$ and  $U=0.14$ ($\omega_{p}=0.1$ and $\omega_{b}=0.48$). See the 
projection of the density maximum showing the bipolaron motion. }
\end{center}\end{figure}

A strong  softening of the pinning mode of the bipolaron is usually associated with a
high mobility. This can be tested by integrating the classical dynamical 
 equations of the  lattice with the effective potential $F_{ad}(\{u_i\})$
 taking as initial condition the bipolaron configuration with a small 
 perturbation in the direction of the pinning mode (see a similar 
 calculation for a moving breather in ref.\cite{CTA96}).
 The moving bipolaron shown fig.\ref{fig4} corresponds to the dip of 
 the pinning mode of fig.\ref{fig3}. 
 Although it is often believed that a high mobility for a localized object 
 in discrete lattice can be achieved only when the size of this object is large 
 compared to the lattice spacing, this moving bipolaron shown is 
localized on few sites only and also still strongly bounded (see inserts of 
fig.\ref{fig3}). It is  nevertheless highly mobile. 
During the motion, it can be viewed as exchanging from state (S0) to 
(S1) and vice-versa.

Relatively small perturbation of the model parameters suffices to 
raise the frequency of the pinning mode and then, it can be checked that 
the quality of the bipolaron mobility diminishes and progresively disappears. 

\section{Bipolaronic Ground-states in Two dimensions}

In the  2D model, a smaller size (up to $14 \times 14$ ) for the lattice 
turns out to be sufficient for accurate calculations because the polaronic
states remains quite localized while they exist. 

\begin{figure} \begin{center}
\includegraphics [width=12cm]{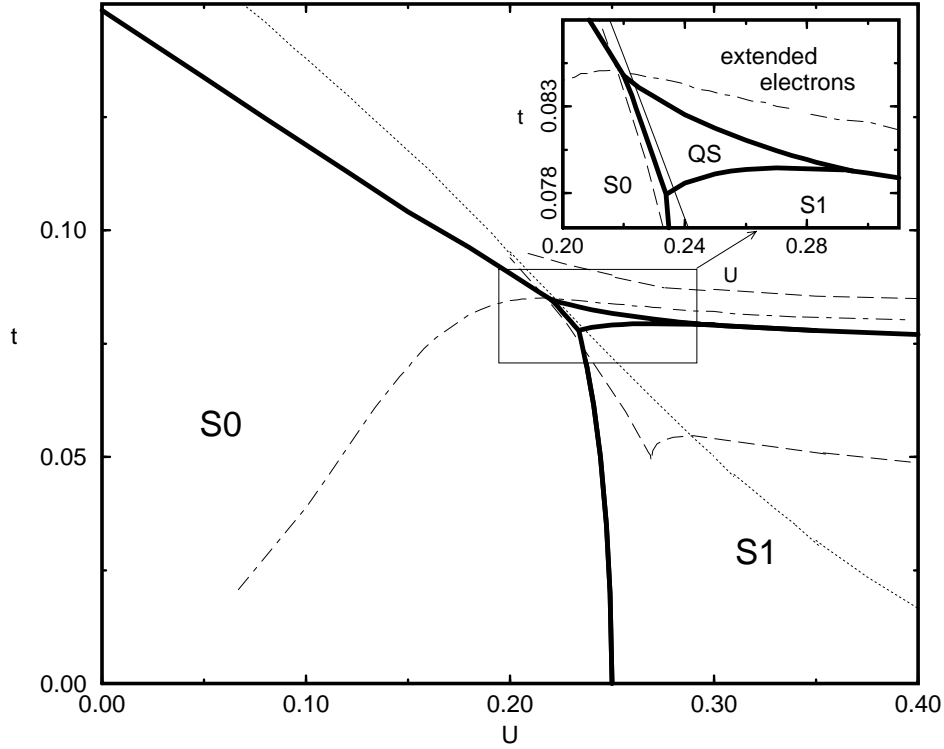}
\caption{\label{fig5}
Phase diagram versus $U$ and $t$ of the bipolaron in the 2D 
Holstein-Hubbard model with four phases separated by first order 
transition lines corresponding to bipolarons (S0), (S1), (QS) and 
extended electrons. The thin dotted line is the limit of metastability of bipolaron(S0), the dot-dashed line is the one of (S1) and the dashed lines are those of (QS). Insert: Magnification of the region of 
fig.\ref{fig5} around the triple point
between phases (S0), (S1) and (QS). }
\end{center}\end{figure}

The obtained phase diagram is shown fig.\ref{fig5}. For $t$ large enough, 
the ground-state corresponds to extended electrons. For $t$ small, there 
is a first order transition as in the 1D model, between bipolarons (S0) and (S1) but 
also in addition, there is a small domain where the bipolaron (QS) 
which was initially unstable for $t$ small, recover its stability and even becomes
the ground-state.

\begin{figure} \begin{center}
\includegraphics[width=12cm]{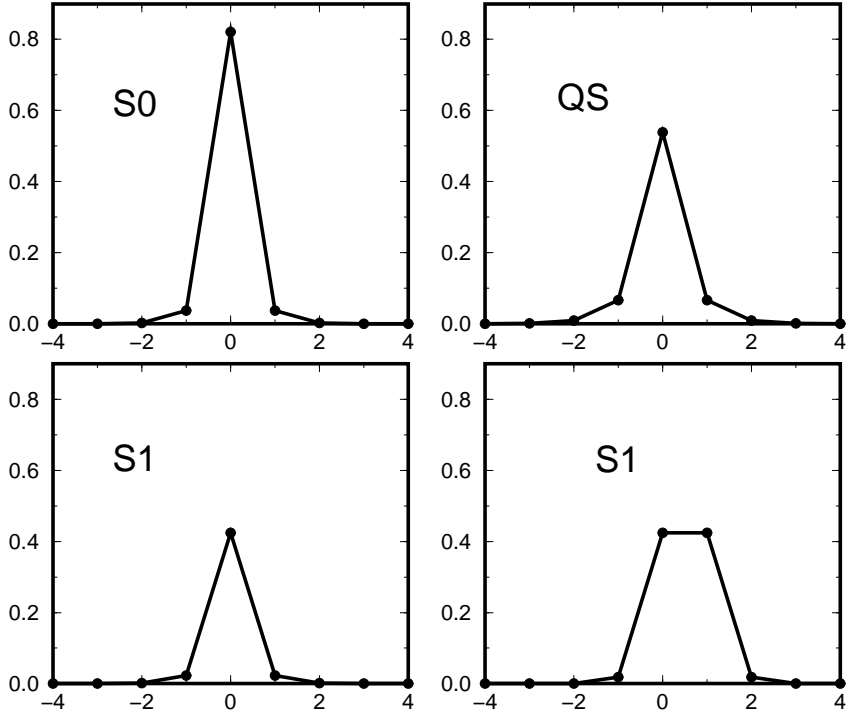}
\caption{\label{fig6}
Electronic density profiles versus site $i$ at the triple
point $t=0.0779$, $U=0.234$ 
for the bipolarons (S0), (QS) along the symmetry $x$ axis 
and the  profiles of (S1) along the symmetry $x$-axis and the 
transverse $y$-axis. These three bipolarons have the same energy.}
\end{center}\end{figure}

The profiles of the three types of coexisting bipolarons (S0), (S1) and 
(QS) with a rather small spatial extension, are shown fig.\ref{fig6}.   

\begin{figure} \begin{center}
\includegraphics[width=12cm]{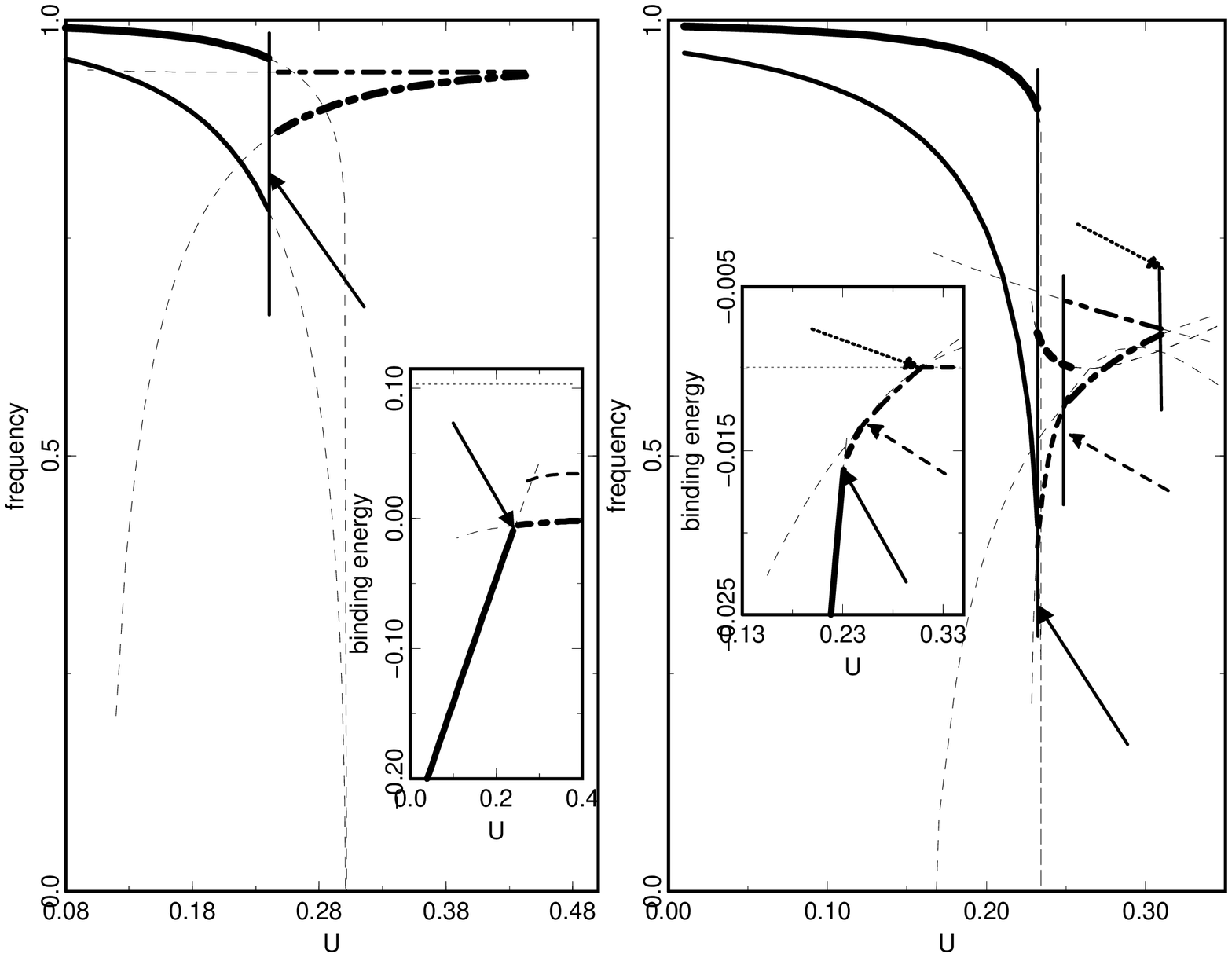}
\caption{\label{fig7}
Frequencies versus $U$ of the pinning modes (thick lines) and the breathing 
modes (thin lines) for bipolarons (S0) (full lines), (S1) (dot-dashed 
lines)and (QS) (dashed lines) at $t=0.05$ (left) and  $t=0.0791$ (right).
The inserts show the corresponding binding energies
for breaking the bipolarons into two polarons far apart. Arrows show the
first order transitions and horizontal line is energy of extended electrons.}
\end{center}\end{figure}

Insert fig. \ref{fig7} shows the energies of the bipolarons on several 
lines at constant $t$. Although, it becomes relatively small, the 
bipolaron binding energies remains non negligible in the vicinity of the triple 
point.

As in 1D, the PN energy barrier is likely depressed in the vicinity of the
first order transition line between (S0) and (S1) or (QS) and (S1)
but  difficult to calculate accurately.
By contrast, the pinning and breathing frequencies of the bipolarons are much 
 easier to calculate and shown fig.\ref{fig7} as a function of $U$
 for several values of $t$. For (S0) and (QS) there are two degenerate pinning modes, one corresponding to the $x$ direction and the other one to the $y$ 
 direction.  Fig.\ref{fig7} shows that there is significant softening of
 both the pinning modes and the breathing mode essentially 
 in the vicinity of the triple point. However, this softening is not sufficient
 for allowing the bipolaron mobility in that region (as confirmed by our tests). 
One may consider that a model involving onsites couplings does not 
favor the bipolaron mobility. Further studies on modified Holstein-Hubbard 
models and particularly models chosen in order to approach a better description
of the cuprates layers $Cu0_{2}$, could perhaps produce classically mobile
bipolarons in $2D$.

\section{Discussion and Concluding Remarks}

Our results are not restricted to the original Holstein-Hubbard model.  We 
already found that the three types of  bipolarons (S0), (S1) and (QS) 
persist in appropriate domain of parameters for modified 
Holstein-Hubbard models for 
example when introducing nearest neighbour Hubbard interaction. We 
believe that few changes in the model which could be physically realistic,
could favor the classical mobility. For example, a phonon dispersion chosen
with an appropriate sign, should increases moderately the spatial extension
of the bipolarons and  thus favor its classical mobility.

In any case, although the classical mobility of a bipolaron should 
favor its quantum mobility,  this condition is not necessary.
When the quantum lattice fluctuation are taken into account ( $\gamma \neq 0$ 
in eq.\ref{hamiltreduc}, the quantum correction to the bipolaronic 
solutions should involve at the lowest order  a RPA term 
(corresponding to small local quantum
fluctuations) and the most important term concerning its physical 
consequences, a tunnelling term raising the spatial degeneracy of 
the bipolarons. The bipolarons should thus form bands.
 
Approximate methods for calculating these bands width analogous to those 
developped in ref.\cite{AQ89,QCRA93} can be 
used.  In the vicinity of the triple point of the phase diagram \ref{fig6},
the three types of 
bipolarons (S0), (S1), (QS) have the same energies and are resonant.
There are four hybridized bands associated with a sharp increase of 
each bandwidth with relatively small gaps for moderately small 
$\gamma$ which makes globally a broad band while elsewhere the 
quantum mobility of the bipolarons sharply diminish.
This current work shall be described in a forthcoming 
publication \cite{PA97}. 

Although we only considered two electrons in the lattice, we could 
expect reasonably that a high bipolaron mobility could persist at 
finite density provided it remains sufficiently far from the half 
filled band. Approaching this limit, the hard core interaction should 
prevent this mobility and produce a frozen structure of 
polaron (likely a SDW or an antiferromagnet).

Before ending this paper, let us briefly check the physical relevance of our results.
These preliminary studies indicates that although relatively weak, the binding
energies of the bipolarons (for producing two extended polarons) in the
vicinity of the triple point are about $0.005$ with our energy unit $E_{0}$.
This chemical energy has to be measured in eV as well as the 
Hubbard term (which is about 8eV for the ion $Cu^{++}$). Thus it is not 
unreasonable that this binding energy range within several $10^{-1}$ eV
which is comparable to the critical temperature energies of the
cuprates. 

In summary, we found that in the 1D model when $U$ increases from zero, the
bipolarons are pairs of polarons at distance $0$ (onsite bipolaron (S0), at distance $1$ (two-sites 
bipolaron or Spin Resonant Bipolaron) (and further but with a binding 
energy going to zero very fast) and that there is first order 
transitions between these different configurations. In the 2D model, 
we find again the onsite bipolarons (S0) and the 2-site bipolaron (S1) 
but a new kind of bipolaron called quadrisinglet. There is a triple 
point in the phase diagram where the three kinds of bipolarons have 
the same energy. 
There is a significant phonon softening at the first order transition. 
In 1D, this softening can become almost complete at moderately large 
values of the transfer integral. Then, the classical bipolaron mobility has
 been effectively observed by direct simulation. 
By contrast, in 2D,  the phonon softening has not been found sufficient 
to produce the classical mobility, but nevertheless one can expect 
a sharp effect of the bipolaron resonance at the triple point, when 
the quantum lattice fluctuations are taken into account. 

We also bring more arguments which maintain our early conjectures 
\cite{Aub92,Aub95}that 
the possible origin of 
high $T_{c}$ superconductors originate from an exceptional combination 
of circonstances consisting in a well-balanced competition between 
strong repulsive electron-electron interactions and a strong 
electron-phonon interactions.

\end{document}